\documentstyle[11pt,newpasp,twoside,epsf]{article}
\def\edcomment#1{\iffalse\marginpar{\raggedright\sl#1\/}\else\relax\fi}
\reversemarginpar

\begin{document}
\title{Gravitational lensing by globular clusters and Arp objects}

\author{Alexander Yushchenko}
\affil{ Odessa Astronomical observatory, Odessa National University, park Shevchenko,
        Odessa, 65014, Ukraine}
\affil{Department of Earth Science Education, Chonbuk National University, 
        Chonju, 561-756, (South) Korea }
\author{Chulhee Kim}
\affil {Department of Earth Science Education, Chonbuk National University,
        Chonju, 561-756, (South) Korea, }
\author{Alexander Sergeev}
\affil{The International Centre for Astronomical, Medical and Ecological
       Research of the Russian Academy of Sciences and
       the National Academy of Sciences of Ukraine,
       Golosiiv, Kiev, 03680, Ukraine  }
\author{Panos Niarchos, Vasilis Manimanis}
\affil{Department of Astrophysics, Astronomy and Mechanics, Faculty of Physics,
        University of Athens, 157 84 Zografos, Athens, Greece}

\begin{abstract}
 We try to explain quasar-galaxy associations by gravitational lensing
 by globular clusters, located in the halos of foreground galaxies.
 We propose  observational test for verification of this hypothesis.
 We processed SUPERCOSMOS sky survey and found
 overdensities of star-like sources with zero proper motions in the
 vicinities of foreground galaxies from CfA3 catalog.
 We show mean effect for galaxies with different redshifts.
 Two effects can explain observational data -
 these are lensing by globular clusters and lensing by dwarf galaxies.
 We made CCD 3-color photometry with 2.0-1.2 meter telescopes
 to select extremely lensed objects around several galaxies
 for spectroscopic observations.
\end{abstract}

\section{Introduction}

 35 years ago Arp (1968) discovered quasar-galaxy associations.
 Among the explanations of these phenomena was gravitational lensing,
 but for a long time the objects  which can act as
 gravitational lenses were unknown.
 Baryshev, Raikov, \& Yushchenko (1993) and
 Yushchenko, Baryshev, \& Raikov (1998) pointed,
 that  stellar globular clusters (GC) and
 dwarf galaxies  in the vicinities of foreground galaxies (FG)
 can be these gravitational lenses. The cores of Seyfert and other
 active galaxies can be the background sources (BS).
 Yushchenko et al. (1998) developed the software
 for calculations the amplifications of extended sources by King objects.
 The amplification can reach 5-10 magnitudes in typical cases.
 Baryshev \& Ezova (1997) showed, that in the case of fractal Universe,
 gravitational lensing can explaine observational data.
 Bukhmastova (2001, 2002) made new catalog of QSO-galaxy associations,
 proposed several new tests and estimated the influence
 of gravitational lensing on the luminosity function of BS.

 Yushchenko (1999) proposed simple observational test for validation
 of this hypothesis. If Seyfert galaxies after amplifications look like QSO,
 than we must observe ordinary galaxies amplificated by GC. The number
 of ordinary background  galaxies is two orders higher than that of
 active galaxies.
 And the mean number of GC in the halos of FGs is near 100.
 Yushchenko (1999, 2000) pointed that the surface density
 of star-like sources around FGs can be  used as a test and found
 that this effect is detectable.
 Yushchenko et al. (2001) published the first results
 of processing the SUPERCOSMOS survey (Hambly et al., 2001). The nature appears
 more beautiful than we can predict - we found overdensity and underdensity
 of possible extremely lensed objects (ELO) around FGs  with different
 redshifts.

\section{Results from SUPERCOSMOS survey}

 SUPERCOSMOS is a digitization
 of photographic survey of the southern hemisphere observed with
 Schmidt telescopes at two different epochs.
 The limiting magnitude is near 21. We selected from the survey
 only star-like images with zero proper motions as a background images.
 We found the number densities of these objects around
 19413~  FGs from CfA3 catalog (Huchra et al., 1995) and around
 the dummy centers without galaxies in  concentric rings
 (the widths of the rings were 30 arcseconds).
 Random centers showed zero effect.
 We expected to find overdensities around FGs.
 But the results were different for different groups of galaxies.
 On  Fig. 1a,~b one can see the typical results for groups of FGs with
 different redshifts.

 The expected overdensity is well detected for  FGs
  with higher redshifts.
 For  FGs with  smaller redshifts we can see underdensity.

 The integration of the square between the line of number densities
 for dummy centers and the line for galaxies give us the number of
 objects which produce the overdensity or underdensity. On the Fig. 1c
 we show the plot of the number of objects responsible for underdensity
 or overdensity
 for different groups of FGs against the redshifts.

 On the Fig. 1d we show the projected linear radii in the halo
 of FG where underdensity or overdensity
 reachs 2/3 of it's final value for different groups of galaxies. We omitted the
 groups with the lowest and highest redshifts - the results for these groups
 are very uncertain.
 
\begin{figure}
\plotone{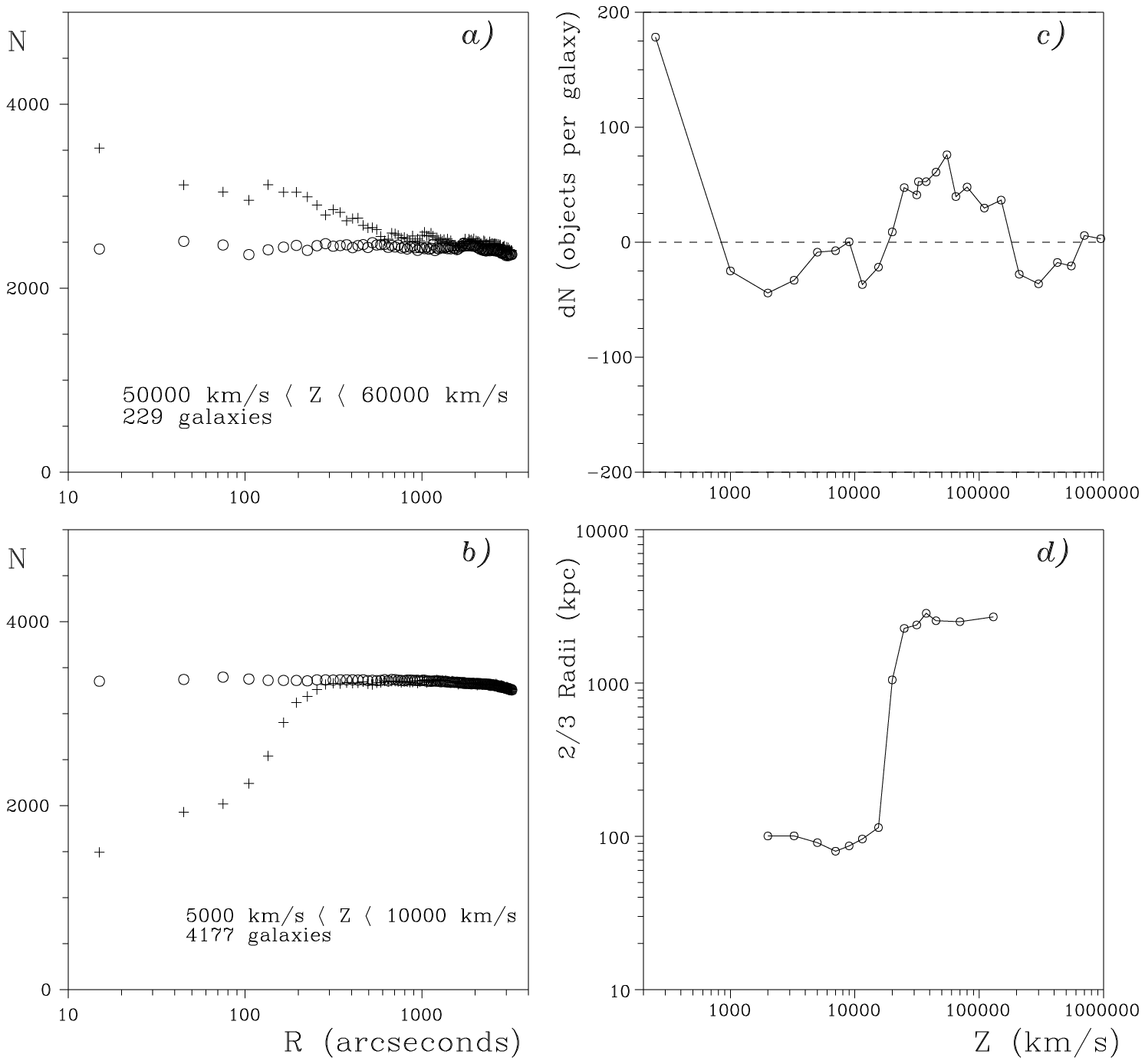}
\caption{Results of processing the SUPERCOSMOS survey. \\
   $a)$ The mean surface density of star-like sources with zero proper motions
        in the vicinities of 229 galaxies with redshifts
        from 50000 to 60000 km/s.
        The axes are the distance from the galaxy center in arcseconds
        and the number density of investigated objects per square degree.
        Circles and crosses - mean results for dummy centers without
        galaxies, and for centers with galaxies respectively. \\
   $b)$ The same as $a)$ but for 4177 galaxies with redshifts
        from 5000 to 10000 km/s. \\
   $c)$ The mean number of objects, which produce overdensity
        or underdensity for groups of galaxies with different redshifts.
        The axes are the redshift in km/s and the number of objects,
        responsible for overdensity or underdensity. \\
   $d)$ The value of linear radii, were overdensity or underdensity of
        objects reachs 2/3 of its final value (if integrate from the zero radii)
        for  groups of galaxies with different redshifts.
        The axes are the redshift in km/s and the linear radii in
        kiloparsecs. }
\end{figure}

\section{Conclusion}
 
  We found the number densities of point-like objects
  with zero proper motions
  in the vicinities of FGs.
  The  overdensity is observed around FGs with Z=0.08-0.5.
  The FGs with Z$<0.08$ show the underdensity.
  The projected linear radii where the effect reachs 2/3 of it's final value
  is different for groups were we can observe underdensity or overdensity
  of investigated objects.
  The projected linear radii is near
   100 kpc in the groups with underdensity and near 2500 kpc in the groups
  with overdensity.

  Yushchenko et al. (2001) pointed, that it is necessary to take into
  account the microlensing. We have the software which
  permit us to simulate the propagation of the light beams through
  the GC.  The results of calculations show that we can explain the
  underdensity for the nearest groups of FGs.
  The GC in this case must contain a big population of
  planets in the outer zones of GC.
  The radii near 100 kps is the radii of the halo of mean galaxy
  with GC in this halo.

  The value of radii for more distant FGs - 2500 kpc, can
  be explained by lensing effect produced by dwarf spheroidal galaxies.
  The number of dSph galaxies in the Local Group is comparable with
  the value of overdensity around the galaxies with Z=0.08-0.5.
  dSph galaxies and GC
  in the small groups of galaxies, like Local Group, can explain
  the observational effect.

   For final verification of our hypothesis it is necessary to point ELOs
   and to measure their redshifts.
   We made 3-color photometry in the vicinities of several galaxies with
   2 m telescope located at peak Terskol (Ukraine-Russia),
   1.8 m telescope of Bohyunsan observatory (South Korea) and 1.2 m
   telescope of Kryonerium observatory (Greece).
   The list of objects for spectroscopic observations
   is in preparation now. The used methodics were described by Yushchenko,
   Niarchos, \& Manimanis (2001)

 This work was supported by the grant of Post-Doc Program, Chonbuk National 
 University (2002)

\end{document}